\documentstyle [12pt] {article}
\begin{document}
\centerline{\Large \bf Nonsingular vortices-skyrmions for odd Landau}
\centerline{\Large \bf level fillings in 2d system}
\centerline{$\quad $}
\centerline{S.V.Iordanskii, S.G.Plyasunov}
\centerline{\small \sl{L.D.Landau Institute for Theoretical Physics Russian Ac.Sci.}}
\centerline{\small \sl{142432 Chernogolovka Moscow district Russia}}

\vskip -13mm

\begin{abstract}
Using gradient expansion method the particle number, the energy and the action
were calculated at the formation of a vortex-skyrmion.Contrary to other papers
on this subject no approximation of single Landau level was used.
Taking into account the nearest Landau levels essentially changes the
expression for the vortex energy and gives some simple physical interpretation
of the main results. It is shown that the formation of a vortex skyrmion
gives the gain in the thermodynamic energy and therefore they must be
spontaneously created near the odd fillings. The Hopf term in the skyrmion
action is calculated and corresponds to a fermionic picture.
\end{abstract}

A large number of theoretical works is devoted to the consideration of
macroscopical  skyrmioniclike excitations near odd Landau level fillings in
2d systems under Quantum Hall Effect conditions. In the work [1] the
phenomenological approach of the Chern-Simons theory was used. It was shown the
existence of such exitations and their energy was calculated. In [2] the
energy was calculated using Hartree-Fock approximation for single Landau level
projected wave functions. The gradient expansion method was used in [3]
for the same model assuming the large size of a skyrmion. The energy and the particle
number were found analitically. These results were confirmed in [4] and the
general technique was developed for the calculation of arbitrary terms in
gradient expansion. In the model of projected functions the skyrmion action
was calculated [5] with the topological term (see the proper discussion [6]).
The experiments give some evidence of skyrmion formation [7] but some
experimentalists cast some doubts concerning these results[8].

Single Landau level projected functions approximation is usually justified
by the large value of the cyclotron energy $\hbar\omega_c$ compare to Coulomb
interaction $e^2/l_H$ where the magnetic length $l_H^2=c\hbar/eH$ and $H$ is the
external magnetic field. The appropriate calculations are rather elaborate and contain
some additional assumptions. The final expressions for the electron energy and
the density are obtained by cumbrsome calculations without any direct physical
interpretation.In the present work we show that the approximation of the
projected wave functions is not sufficient for the adequate skyrmion description.
Some preliminary results were published in [9] and the present paper is
more complete with the justification of the results by gradient expansion
method.

Skyrmions correspond to the rotation of the second quantized operators-spinors
using nonuniform rotation matrix
$U(\vec{r})$. In that case the initial spinors in the laborator frame $\psi$
are expressed in terms of the new spinors $\chi$ with spin up in the local
frame according to
the relation $\psi(\vec{r})=U(\vec{r})\chi(\vec{r})$.The matrix $U(\vec{r})$
can be parametrized by three Euler angles
\begin{displaymath}
\nonumber
U(\vec{r})=U_z (\gamma (\vec{r})) U_y
(\beta (\vec{r})) U_z (\alpha (\vec{r}))
\end{displaymath}
\begin{displaymath}
U_z (\alpha)=\cos \frac{\alpha}{2}+i\sin \frac{\alpha}{2}\sigma_z
\end{displaymath}
\begin{displaymath}
U_y (\beta)=\cos \frac{\beta}{2}+i\sin \frac{\beta}{2}\sigma_y ,
\end{displaymath}
where ${\sigma_x, \sigma_y, \sigma_z}$ are Pauli matrices. At large distances
from the core for finite value of g-factor the average spin must be oriented
along the external magnetic field to have a finite value of the total energy.
Therefore the angle $\beta$, counted from the magnetic field direction must
rapidly vanish for $r\rightarrow\infty$  (it can be shown exponentially)
We assume that the matrix $U(\vec{r})$ has no singularities at any $\vec{r}$.That
corresponds to a nonsingular behaviour of the matrices
\begin{displaymath}
\nonumber
A_k=-iU^+\frac{\partial U}{\partial x_k}=\Omega^l_k(\vec{r})\sigma_l
\end{displaymath}
where $k=x,y$ and $l=x,y,z$, Pauli matrices
$$
\sigma_z=\left(\begin{array}{cc}
1&0\\0&-1
\end{array}\right)
\quad\sigma_x=\left(\begin{array}{cc}0 &1\\1&0\end{array}
\right)
\quad\sigma_y=\left(\begin{array}{cc}0 &{-i}\\{i}&0\end{array}
\right)
$$

The expression for the functions $\Omega_k^l$ are easily obtained by the direct
differentiation
of $U$
\begin{displaymath}
\Omega_k^z=\frac{1}{2}(\partial_k{\alpha}+cos{\beta}\partial_k\gamma)
\end{displaymath}
\begin{equation}
\label{b}
\Omega_k^x=\frac{1}{2}(\sin{\beta}\cos{\alpha}\partial_k\gamma-\
sin{\alpha}\partial_k\beta)
\end{equation}
\begin{displaymath}
\Omega_k^y=\frac{1}{2}(\cos{\alpha}\partial_k\beta+\sin{\beta}\sin{\alpha}\partial_k\gamma)
\end{displaymath}
The non trivial topology defined by the matrix  $U(\vec{r})$, is connected
with the
properties of the mappings $\alpha(\vec{r}),\gamma(\vec{r})$, where $\vec{r}$
is varied along a circumference of large radii. The degree of the mapping of 2d
plain $x,y$ on the sphere parametrized by the angles $\gamma$ and $\beta$
coincides with the degree of the mapping of the circumference on the
circumference i.e. the winding number for the vortex singularity of
$\gamma(\vec{r})$. For a nonsingular $\Omega_k^l(\vec{r})$ it is necessary
to have coinciding singularities in $\gamma(\vec{r})$ and $\alpha(\vec{r})$ at the
the point where $\cos{\beta}=-1$. Therefore in the matrix $U$ must be represented
all three Euler angles and the corresponding spinor $\psi(\vec{r})$ has vortex
singularity of it's phase at large distances with the integer winding number
and is simple in 2d plain. For that reason it is more appropriate to talk about
a non singular vortex with skyrmionic core by the anology with $He_3$. The winding
numbers are here any integgers contrary to the case of $He_3$ where they are
even [10]. The integral
$$
\frac{1}{2\pi}\int rot\vec\Omega^zd^2r=Q
$$
is a topological invariant and is expressed directly through the change of the
phase for the spinor $\psi$ along a large circumference.

The pair interaction Lagrangian for electrons :
\begin{displaymath}
L=\int\left[i\psi^{+}\frac{\partial\psi}{\partial t}-
\frac{1}{2m}\psi^{+}(-i\partial_k-A_{0k})^2\psi\right]d^2rdt+\\
\end{displaymath}
\begin{equation}
\frac{1}{2}\int V(\vec{r-r'})\psi^{+}(\vec r)
\psi^{+}(\vec{r'})\psi(\vec{r'})\psi(\vec{r})d^{2}r'd^2rdt
-gH\int \psi^{+}\sigma_z\psi d^2rdt
\end{equation}
transforms by the matrix $U$ into the Lagrangian in terms of spinors $\chi$
\begin{displaymath}
L'=\int i\chi^{+}\left[\frac{\partial\chi}{\partial t}-
\Omega^l_t\sigma_l\chi-
\frac{1}{2m}(i\partial_k-A_{0k}+\Omega_k^l\sigma_l)^2\chi\right]d^2rdt
\end{displaymath}
\begin{equation}
\label{c}
-\frac{1}{2}\int V(\vec{r-r'})
\chi^{+}(\vec r)\chi^{+}(\vec r')\chi(\vec r')\chi(\vec r)d^2r'd^2rdt
+
gH\int U^{+}\sigma_zU)\chi d^2rdt
\end{equation}
without any approximation. We assume that matrix $U$ depends on the time also.
The quantity $\Omega_t^l$ corresponds to (\ref{b}) but with the time derivative.
$A_0$ is the vector potential of the external uniform magnetic field. The
Lagrangian (\ref{c}) can be easily obtained by direct differentiation
taking into account some identity obtained by the differentiation of the
of the identity $U^{+}U=1$.

The size of the vortex-skyrmion core is defined by Coulomb interaction which
tend to  increase the domain with large change of the density and by Zeeman
energy which tend to diminish this domain due to the positive energy of the
spins non aligned to the external magnetic field. At large distances Coulomb
is small and only Zeeman energy and the energy connected with spin direction
gradients are essential. That gives the exponential decay of non aligned spins.
For a small value of the g-factor the size of the core will be large
and the derivatives of the matrix $U$ become small that justify the use of the
gradient expansion for the calculations. We shall treat this case interesting
mostly in topologically invariant terms not dependig on deformations of $U$.
That allow us to concider no Zeeman energy or taking it into account only in the
first order of the perturbation theory.

The effective action depending on the matrix $U$ is defined by the integration
out the fermions in the correspondent expression for the statistical sum and
has the form [11] $S=iTr\ln G$, where $G$ is electron Green function.
The Trace is over whole set of variables including spin and space-time.This
action will be calculated by the gradient expansion in $U$ using Hartree-Fock
approximation which we justify later.

\section{ Green function in Hartree-Fock approximation}

We use Hartree-Fock approximation for the calculation of Green function
which is valid for fully filled Landau level with the accuracy of the order
$\textstyle V_{int}/ \textstyle \hbar\omega_c$ assumed to be small.Such an
approach allows to write the electronic Hamiltonian as
\begin{displaymath}
H=\int\chi^{+}\left[\Omega^l_t\sigma_l+\frac{1}{2m}(-i\partial_k-A_{0k}+
\Omega_k^l \sigma_l)^2\right]\chi d^2r+
\end{displaymath}
\begin{equation}
\label{d}
\int V(\vec{r-r'})<\chi^{+}(\vec r')\chi(\vec r')>
\chi^{+}(\vec r)\chi(\vec r)d^2r'd^2r-
\end{equation}
\begin{displaymath}
\int V(\vec{r-r'})
<\chi^{+}(\vec r')\sigma_l\chi(\vec r)>
\chi^{+}(\vec r)\sigma_l\chi(\vec r')d^2rd^2r'
\end{displaymath}
The interaction term becomes the sum of exchange and direct interaction and
small Zeeman energy is ommited. In the direct unteraction we must ommit zero
Fourrier component for Coulomb case due to compensating uniform charge
(see e.g.[11]).

We assume that all the average quantities are close to their uniform values for
fully filled Landau level. We restrict our consideration by the model of the local
exchange neglecting the difference $\vec r-\vec r'$ that formally corresponds
to the interaction distance in $V(r)$ small compared to the magnetic length.
Probably the transition to the local exchange is quite unessential and
topological terms does not
depend on it but the use of such an approximation greatly simplifies the
calculations. The final results are easily generalized on the case of the
nonlocal exchange. Therefore we shall deal with the hamiltonian
\begin{equation}
\label{e}
H=\int \left[\chi^{+}\Omega_t^l\sigma_l\chi^{+}+\frac{1}{2m}\chi^{+}(-i\partial_k-
A_{0k}+\Omega_k^l\sigma_l)^2\chi-\gamma\rho\chi^{+}\sigma_ln_l\chi^{+}+
V_0\rho\chi^{+}\chi\right]d^2r
\end{equation}
where $\gamma$ is the exchange constant and $V_0$  is the constant of direct
interaction wich we shall omit as in the case of Coulomb interaction, $\rho$
is the average density. We suppose that $\chi$ spinors are oriented in z-direction
coinciding
with the average local spin direction in the laborator frame. Assuming a small
value of the angle velocities ${\Omega^l_\mu}$ and their derivatives
$\partial\Omega\sim\Omega^2$ we shall calculate Green function by the perturbation
theory separating the hamiltonian $H=H_0+H_1+H_2$ into three parts
\begin{equation}
\label{f}
H_0=\int\chi^{+}\left[\frac{1}{2m}(-i\partial_k-A_{0k})^2-\gamma\rho\sigma_l
\vec{n_l}-\mu\right]\chi d^2r
\end{equation}
\begin{equation}
\label{f1}
H_1=\frac{1}{m}\int \chi^{+}[\Omega_{k}^{l}\sigma_{l}(-i\partial_{k}-A_{0k})
+\Omega^{l}_{t}\sigma_l]\chi d^2r
\end{equation}
\begin{equation}
\label{f2}
H_2=\frac{1}{2m}\int \chi^{+}\left[(\Omega_k^l\sigma_l)^2-i\frac
{\partial{\Omega_k^l}}{\partial r_k}\sigma_l\right]\chi d^2r
\end{equation}
We use grand canonical ensemble introducing chemical potential $\mu$.

Green function for the unperturbed Hamiltonian is Green function for
noninteracting electrons in uniform magnetic field. We assume that at the last
filled Landau level only the lowest spin sublevel is occupied. For simplicity
we consider here only the case when Landau index $s=0$ when unperturbed
Green function acquires the form
\begin{equation}
\label{g1}
G_0(\vec r,\vec r',t,t')=-i<T\chi(\vec{r},t)\chi^{+}(\vec{r}',t')>=
\sum_{p,s}\int g_s(\omega)e^{i\omega(t'-t)}
\Phi_{sp}(\vec r)\Phi_{sp}^{*}(\vec r')\frac{d\omega}{2\pi}
\end{equation}
Here $T$ is a symbol of time product for fermions, the sum is over all $s$ and
$p$, spin matrices $g_s(\omega)$ corresponds to fully occupied Landau level
$s=0$ with spin up in the local frame. Other states are empty. All quantities
are measured in units where $H_0=1,l_H=1$ and $\hbar=1$. The normalized wave
functions $\Phi_{sp}(\vec r)$ corresponds to eigenfunctions in Landau gauge.

It is easy to find matrices $g_s(\omega)$ for the hamiltonian (\ref{f})
\begin{equation}
\label{g2}
g_0(\omega)=\frac{1}{\omega+(\gamma\rho-i\delta)\sigma_z+\mu}
\end{equation}
\begin{equation}
\label{g3}
g_s(\omega)=\frac{1}{\omega +\gamma\rho\sigma_z-\frac{s}{m}+\mu+i\delta}
\end{equation}
where $\delta\rightarrow(+0)$, the energy of zero Landau level $1/{2m}$ is
included in chemical potential.

Using the expression (\ref{g1}) we can develop the perturbation theory
expansion for Green function of the total hamiltonian $G=G_0+G_1+G_2+...$. The
corresponding diagrams are shown in fig 1 up to the third order.

It is usefull for further calculations to represent the operator entering
(\ref{f1}) in the form
\begin{equation}
\label{g4}
\frac{1}{m}\Omega_k^l\sigma_l(-i\partial_{k}-A_{0k})=\frac{1}{m}(\Omega_{+}^l\pi^{-}+
\Omega_{+}^l\pi^{+})\sigma_l
\end{equation}
The operator $\pi^{+}\Phi_{sp}=\sqrt{2(s+1)}\Phi_{(s+1)p}$ increases Landau
index,and the operator $\pi^{-}\Phi_{sp}=\sqrt{2s}\Phi_{(s-1)p}$ decreases
Landau index what follows directly from the properties of oscillator wave
functions and
\begin{equation}
\label{g5}
\Omega_{+}^l=-\frac{i\Omega^l_x+\Omega^l_y}{2}, \\
\Omega_{-}^l=\frac{i\Omega^l_x-\Omega^l_y}{2}
\end{equation}.

The correction of the first order in  $\Omega$ to Green function is
\begin{displaymath}
G_1(\vec r,\vec r',t,t')=\int e^{-i\omega(t-t')}
g_s(\omega)\Phi_{s,p}(\vec r)\Phi_{s,p}^{*}(\vec r_1)\nonumber\\
\end{displaymath}
\begin{displaymath}
\times[\frac{1}{m}(\Omega_{+}^l(\vec r_1)\pi^{-}+
\Omega_{-}^l(\vec r_1)\pi^{+})+\Omega_{t}^l(\vec r_1)]\sigma_lg_{s'}
(\omega')\nonumber\\
\end{displaymath}
\begin{equation}
\label{g6}
\times e^{-i\omega'(t_1-t')}\Phi_{s'p'}
(\vec r_1)\Phi_{s'p'}^{*}(\vec r')\frac{d\omega}{2\pi}\frac{d\omega_1}{2\pi} d^2r_1dt_1
\end{equation}
We are interested in Green function $G(r,r;t,t+\delta)$ defining the density.
In that case the only nonvanishing terms are with  $s=0,s'=1$ , $s=1,s'=0$
and $s=s'=0$.
The others vanish because of analytical properties of $g_s$ for $s>0$ or have
some additional derivatives. It is possible to neglect the difference $t_1-t$
for the same reason. We do not consider at first the term with
 $\Omega_t^l$ which can be easily calculated. Integrand in (\ref{g6}) is
vanishing rapidly when $| \vec r-\vec r_1|$ is increased above magnetic
length. Therefore it is possible to expand $\Omega(\vec r_1)$ in terms of
$\vec r-\vec r_1$ up to linear terms. Performing the summation in  $p$ we get
the following expression
\begin{displaymath}
G_1(\vec r,\vec r',t,t+\delta)=\frac{\sqrt{2}}{m}\int g_0\sigma_lg_0e^{i\omega\delta}
\frac{\partial\Omega^l_{-}}{\partial r_k}R_{00}(\vec r,\vec r_1)R_{10}(\vec r_1,\vec r)
(r_{1k}-r_k)d^2r\frac{d\omega}{2\pi}+
\end{displaymath}
\begin{displaymath}
\frac{\sqrt{2}}{m}\int g_0(\omega)
\sigma_lg_1e^{i\omega\delta} \frac{d\omega}{2\pi}
\int R_{00}(\vec r,\vec r_1)R_{01}(\vec r_1,\vec r)
(r_1- r)_k\frac{\partial\Omega_{-}^l}{\partial r_k} d^2r_1+
\end{displaymath}
\begin{equation}
\label{g7}
\frac{\sqrt{2}}{m}\int g_1(\omega)\sigma_lg_0(\omega)e^{i\omega\delta}
\frac{d\omega}{2\pi}
\int R_{11}(\vec r,\vec r_1)R_{10}(\vec r_1,\vec r)
(r_1-r)_k\frac{\partial\Omega_{+}^l}{\partial r_k} d^2r_1
\end{equation}
Here we introduce the functions
\begin{eqnarray}
R_{00}(\vec r,\vec r_1)=\frac{1}{2\pi}\exp(-\frac{(\vec r-\vec r_1)^2}{4}
\label{g8}\\
R_{11}(\vec r,\vec r_1)=\frac{1}{2\pi}[1-\frac{(\vec r-\vec r_1)^2}{2}]
\exp{-\frac{(\vec r-\vec r_1)^2}{4}}\label{g9}\\
R_{10}(\vec r_1,\vec r)=\frac{1}{2\pi}\frac{x_1-x-i(y_1-y)}{\sqrt{2}}
\exp{-\frac{(\vec r-\vec r_1)^2}{4}}\label{g10}\\
R_{01}(\vec r_1,\vec r)=\frac{1}{2\pi}\frac{x-x_1-i(y_1-y)}{\sqrt{2}}
\exp{-\frac{(\vec r-\vec r_1)^2}{4}}\label{g11}
\end{eqnarray}
After the integration in $\vec r_1$  the term wiothout the derivatives of
$\Omega$ vanish because the integrand is an odd function. Doing all
integrations in(\ref{g7}) we get using (\ref{g5}), (\ref{g3}), (\ref{g2})
\begin{displaymath}
G_1(\vec r,\vec r,t,t+\delta)=\frac{1}{4m\gamma}
\frac{\sigma_l-\sigma_z\sigma_l\sigma_z}{2}(div\vec\Omega^l-irot\vec\Omega^l)
\end{displaymath}
\begin{displaymath}
\frac{1}{2\pi}
\left[\frac{\sigma_z\sigma_l-\sigma_l\sigma_z}{4}div\vec
\Omega^l-\frac{i}{2\pi}(\frac{(1+\sigma_z)\sigma_l+
\sigma_l(1+\sigma_z)}{4}rot\vec \Omega^l+\right]
\end{displaymath}
here we consider only the main
spin independent contribution to $g_1\sim{m}$. In the term with $\Omega_t^l$ we
must consider only the case $s=s'=0$ others are of the order of $m$ and can be
neglected. Only the terms with the first order poles in $\omega$ are essential
that results in
$$
G_1^{''}=-i\Omega_t^l\frac{\sigma_l-\sigma_z\sigma_l\sigma_z}{8\pi\gamma\rho}
$$
Finally the first order correction to Green function acquires the form
\begin{equation}
\label{g12}
G_1(\vec r,\vec r,t,t+\delta)=\frac{1}{8\pi}\left[
(\sigma_z\sigma_l-\sigma_l\sigma_z)div\vec\Omega^l -i((1+\sigma_z)\sigma_l+
\sigma_l(1+\sigma_z))rot\vec\Omega^l\right]+
\end{equation}
\begin{displaymath}
\frac{1}{4m\gamma}\frac{\sigma_l-\sigma_z\sigma_l\sigma_l}{2}(div\vec\Omega^l-irot\vec\Omega^l)
-i\Omega_t^l\frac{\sigma_l-\sigma_z\sigma_l\sigma_z}{8\pi\gamma}
\end{displaymath}
 For the density
$\rho=-iSpG(\vec r,\vec r,t,t+\delta)$ we obtain
\begin{equation}
\label{g13}
\rho(\vec r,t)=\frac{1}{2\pi}(1-rot\vec \Omega^z)
\end{equation}
This result was firstly obtained in the approximation of the projected wave
functions [3]. In the terms of unprojected wave functions this result has a
simple physical meaning: the lowest Landau level in the local effective field
$H_{eff}=1-rot\vec \Omega^z $ is fully filled and the density coincides with
the local density of states $\rho=1/{2\pi}l^2(H_{eff})$.

 The results for the first order term in $G$ contain the derivative of
$\Omega$ that corresponds to second order in $\Omega$.Therefore it is
necessary to calculate the formally second order term in $G$
which can be written down simbolically as
\begin{displaymath}
G_2(\vec r,\vec r,t,t+\delta)=\frac{1}{m^2}G_0(\Omega_{+}^l\pi^{-}+
\Omega_{-}^l\pi^{+})\sigma_lG_0(\Omega_{+}^{l'}\pi^{-}+
\Omega_{-}^{l'}\pi^{+})\sigma_{l'}G_0+G_0H_2G_0
\end{displaymath}
In the second order we can neglect the additional derivatives of $\Omega$.
Because the last term is formally of the second order we must
take in it the states with $s=0$ with the second order pole in
$\omega$ which vanish after the integration. In the first term there are two
types of contributions with $g_0g_1g_0$ and $g_1g_0g_1$, where $g$ corresponds
to one of the three Green functions in this expression. The integration over
space variables and  $p$ are easily performed and we get
\begin{displaymath}
\label{g14}
G_2=\frac{1}{m^22\pi}\int [g_0\sigma_lg_1\sigma_{l'}g_0\Omega_{+}^l\Omega_{-}^{l'}
+g_1\sigma_lg_0\sigma_{l'}g_1\Omega_{-}^l\Omega_{+}^{l'}]e^{i\omega\delta}\frac{d\omega}{2\pi}d^2rdt
\end{displaymath}
To calculate the second order correction to the density we need trace of this
expression. Transposing the matrices under the trace we get in the first case
the square of diagonal matrix $g_0^2(\omega)$ having the second order pole in
$\omega$. Using the standart expression for the proper integral we obtain
\begin{displaymath}
SpG_2=\frac{1}{2\pi m^2}\int \left[-2g_1^2\Omega_{+}^l\Omega_{-}^l+2g_1^2\Omega_{+}^l\Omega_{-}^l
\right]d^2rdt=0
\end{displaymath}
Therefore the density is given by the expression (\ref{g13}) up to the second
order.

\section{The calculation of the action,the energy and the particle number}

For the calculation of the action   $S=iSp\ln G$ we use the perturbation
theory for Green function and expand the logarithm
\begin{displaymath}
S=iSp[H_1G_0+\frac{1}{2}H_1G_0H_1G_0+H_2G_0+
\frac{1}{2}(H_2G_0H_1G_0+H_1G_0H_2G_0)+\frac{1}{3}H_1G_0H_1G_0H_1G_0+...]
\end{displaymath}
Here we omit nonessential part without the rotation matrix and only terms
up to the third order are retained. In this section we limit our consideration
by the second order terms. The action of the first order is given by
\begin{displaymath}
S_1=iSp \int \Omega_t^l\sigma_lg_0(\omega)e^{i\omega\delta}\frac{d^2rdt}{2\pi}
\frac{d\omega}{2\pi}+
iSp \int \sigma_lg_0(\omega)e^{i\omega\delta}\frac{d\omega}{2\pi}
(\Omega_{+}^l\pi^{-}+\Omega_{-}^l\pi^{+}) \Phi_{0p}(r)\Phi_{0p}^{*}(r) d^2rdt
\end{displaymath}
The second term can be calculated using integration by parts and we get
\begin{equation}
\label{s2}
S_1=-\frac{1}{2\pi}\int \Omega_t^z drdt-
\frac{1}{2\pi}\frac{1}{2m}\int (i div \vec \Omega^z -rot \vec \Omega^z) d^2rdt
\end{equation}

The action of the second order consists of two terms one of which contains
two $H_1$
$H_1$:
\begin{displaymath}
S'_2=\frac{i}{2m^2}Sp\int (\Omega_{-}^l\pi^{+}+
\Omega_{+}^l\pi^{-})\sigma_lg_s(\omega)\Phi_{sp}(\vec r)\Phi_{sp}^{*}(\vec r')
\end{displaymath}
\begin{displaymath}
(\Omega_{-}^{l'}\pi^{+}+\Omega_{+}^{l'}\pi^{+})\sigma_{l'}g_{s'}
(\omega)\Phi_{s'p'}(\vec r')\Phi_{s'p'}^{*}(\vec r)e^{i\omega\delta}
d^2r'd^2rdt\frac{d\omega}{2\pi}
\end{displaymath}
In this expression we make no difference between $\Omega(\vec r,t)$ and
$\Omega(\vec r',t')$ neglecting derivatives of $\Omega$.The only nonvanishing
terms correspond to $s=0,s'=1$ and $s=1,s'=0$.
Using the properties of the operators $\pi^{+},\pi^{-}$ and performing simple
integration it is easy to get the second order action including the second
term with $H_2$
\begin{displaymath}
S_2=\frac{1}{2\pi}\int Sp \frac{(\sigma_{l'}\sigma_{l}(1+\sigma_z)+
\sigma_{l}\sigma_{l'})(1+\sigma_z)}{2}
\Omega_{-}^l\Omega_{+}^{l'} d^2rdt-\frac{1}{2\pi}\int [\frac{(\vec
\Omega^l)^2}{2m} -
\frac{i}{2m}\frac{\partial \Omega_k^z}{\partial r_k}] d^2rdt
\end{displaymath}
The symmetric part with $l=l'$ of the first term is compensated by the
proper part of the second term. The antysimmetric part containes only
$l,l'=x,y$ and can be rewritten using the identity
$rot \vec \Omega^z=2(\Omega_x^x\Omega_y^y-\Omega_y^x\Omega_x^y)$
which is obtained by differentiation using the identity $U^{+}U=1$. Finally we
get the second order action
\begin{equation}
\label{s3}
S_2=\frac{1}{2m}\int rot \vec \Omega^z\frac{d^2rdt}{2\pi}+
i\int div \vec \Omega^z\frac{d^2rdt}{2\pi}
\end{equation}
The total action up to the second order inclyding (\ref{s2}) is
\begin{equation}
\label{s4}
S=\int \Omega_t^z\frac{d^2rdt}{2\pi}+\frac{1}{m}
\int rot\vec \Omega^z\frac{d^2rdt}{2\pi}-
\mu\int rot\vec \Omega^z\frac{d^2rdt}{2\pi}
\end{equation}
Here we add the term with chemical potential.

These results are in full agreement with the authors preliminary publication
[9] and have simple physical interpretation.The electrons locally fill the
the lowest spin sublevel in the local "effective" magnetic field
$H_{eff}=1-rot \vec\Omega^z$. They fill it fully so the electron density
coincides with the local density of states.This circumstanse justifies
the use of Hartree-Fock approximation in the same manner as it does for the
fully filled
Landau level. The "effective" magnetic field is less than the external one
for vortices with positive $Q$. It must be stressed that the gap beween spin
sublevels is defined by the exchange. The electrons with the spin direction
opposite to average spin direction will see the "effective" field
$H_{eff}=1+rot \vec\Omega^z$.

Let us  consider more detailed expression for the energy in a stationary case
taking into account the change in the interaction terms due to the variation of
the density. For the Coulomb interaction one must assume the existence of the
compensating positive background  recquiring to vanish the zero Fourrier
of the interaction. Because of that there is no linear term in the density
variation for direct interaction. It's expansion begins from the second order
\begin{displaymath}
E_{pot}=\frac{1}{2}\int \frac{e^2}{|\vec r-\vec r'|}rot \vec
\Omega^z(\vec r)rot \vec \Omega^z(\vec r')\frac{d^2rd^2r'}{(2\pi)^2}
\end{displaymath}
The exchange energy begins with a linear term in the variation of the density.
For it's calculation one must express the density of the exchange energy as
a function
of the magnetic field for fully filled Landau level
 $-\frac{e^2}{2\pi{l_H^3}}\sqrt{\frac{\pi}{2}}$ differenciate it in magnetic
 field and multiply by the variation in effective magnetic field due to vortex
formation
\begin{displaymath}
E_{ex}=\int \frac{3e^2}{4\sqrt{2\pi}l_H} rot\vec \Omega^z d^2r
\end{displaymath}

Also it is necessary to take into account an energy correction due to
nonuniform direction of the average spin direction
$J\int (\frac{\partial{n^i}}{\partial{r_k}})^2d^2r$ where the quantity
$J=\frac{1}{16\sqrt{2\pi}}\frac{e^2}{l_H}$, (see e.g.[3,4].
and add Zeeman energy.
Finally we get for the change of the thermodynamic energy due to vortex-skyrmion
formation
\begin{displaymath}
\nonumber
F=\delta<H-\mu N>=-\frac{\hbar\omega_c}{2}Q+
\frac{3e^2}{2l_H}\sqrt{\frac{\pi}{2}}Q+\\
\end{displaymath}
\begin{equation}
\label{s5}
\frac{e^2}{2}\int \frac{rot \vec \Omega^z(\vec r)rot \vec
\Omega^z(\vec r')}{(2\pi)^2|\vec r-\vec r'|}
d^2rd^2r'+\int [\frac{1}{2}J(\frac{\partial{n_i}}{\partial{r_k}})^2+
g\vec H\vec n\frac{1}{2\pi{l_H^2}}]{d^2r}
\end{equation}
We assume here the value of chemical potential
$\mu=\frac{\hbar\omega_c}{2}$
corresponding to the fully filled lowest Landau spinsublevel at large distances
from the core. By assumption the cyclotron is large compare to the interaction
energy and consequently  the quantity $F$ is negative for positive topological
number $Q$ defining the change of the total number of electrons $\delta N=-Q$
at vortex formation. Therefore the formation of nonsingular vortices-skyrmions
give the gain in the thermodynamic energy and they must be spontaneously
created.

 It is possible to calculate also the energy of one particle exitations.
This can be done by the investigation of the poles for $G(\vec r,\vec r,\omega)$.
However it is more simple to use the fact that these energies corresponds to
to the variational derivatives of the total energy over the local density of
the electrons with the different spin direction. It is possible to write down
the
quantity $F$ as a functional depending on the densities
$\rho_{+},\rho_{-}$ where signs denote the direction along and opposite to the
average local spin
\begin{displaymath}
\nonumber
F=\int [\frac{\hbar\omega_c(H_{eff}^{+})\rho_{+}+
\hbar\omega_c(H_{eff}^{-})\rho_{-}}{2}-\mu(\rho_{+}+\rho_{-})-
\frac{1}{2}(\rho_{+}-\rho_{-})^2(\gamma'-
J'(\frac{\partial{n^l}}{\partial{r_k}})^2]d^2r+\\
\end{displaymath}
\begin{equation}
\nonumber
\int \frac{e^2}{2|\vec r-\vec r'|}(\delta\rho_{+}(\vec r+\delta\rho_{-}(\vec r))
(\delta\rho_{+}(\vec r')+\delta\rho_{-}(\vec r'))d^2rd^2r'+
\int g\vec H\vec n(\rho_{+}-\rho{-})d^2r
\end{equation}
In this expression $\gamma'\rho_{+}^2=\frac {e^2}{2\pi{l_H}^3}\sqrt{\frac{\pi}{2}}
-\frac{3e^2}{4\pi l_H}\sqrt{\frac{\pi}{2}}rot\vec \Omega^z$ and $J'\rho_{+}^2=J$.
Variation in densities $\rho_{+},\rho_{-}$ gives us the energy of the hole in
the filled states
\begin{equation}
\label{s6}
\epsilon_{h}=-\frac{\delta F}{\delta\rho_{+}}=
\frac{2e^2}{l_H}\sqrt{\frac{\pi}{2}}(1-\frac{3}{2}l_H^2
rot\vec \Omega^z)+
\end{equation}
\begin{displaymath}
\frac{\hbar\omega_c}{2}rot \vec \Omega^z+\int
\frac{e^2}{|\vec r-\vec r'|}rot \Omega^z(\vec r')\frac{d^2r'}{2\pi}
-J4\pi l_H^2(\frac{\partial{n^l}}{\partial{r_k}})^2-g\vec H\vec n
\end{displaymath}
and the energy of the electron with the locally reversed spin
\begin{equation}
\label{s7}
\epsilon_e=\frac{2e^2}{ l_H}\sqrt{\frac{\pi}{2}}
(1-\frac{3l_H^2}{2}rot \vec \Omega^z)+\\
\frac{\hbar\omega_c}{2}rot \vec\Omega^z-\int
\frac{e^2}{|\vec r-\vec r'|}rot \vec \Omega^z(\vec r')\frac{d^2r'}{2\pi}
+J4\pi l_H^2(\frac{\partial{n^l}}{\partial{r_k}})^2-g\vec H\vec n
\end{equation}

All previous results are obtained by using the perturbation theory in  $\Omega$.
 That means that one particle energy corrections due to $\Omega$ must be small
compared to the initial gap separating the lowest spin sublevel from the higher
one. Therefore the largest additional term in one particle energy
${\hbar\omega_c/2}l_H^2rot\vec\Omega^z\sim {\hbar\omega_c/2}l_H^2/L_c^2$
where $L_c$ is the core size must be small compare to the exchange energy
$\sim e^2/l_H$. That gives the inequality $L_c\gg l_H\sqrt{\hbar\omega_c/e^2}$.
The core size is defined by the competition of Coulomb energy $ e^2/L_c$ and
Zeeman
energy $gH\rho L_c^2$ that gives
 $L_c^3\sim (e^2l_H^2)/gH\gg l_H^3[(\hbar\omega_cl_H)/e^2]^{3/2}$.
Therefore for the validity of the obtained results g-factor must be small
enough.
 The addition of the electron with reversed spin increase the thermodynamic
energy due to exchange and decrease it by Coulomb interaction for
positive $Q$ other terms are essentially less due to
small value of $\Omega$. Coulomb energy increases with $Q$ and inverse size of
core whereas exchange energy is local and does not depend on the core structure.
Therefore it is possible to have metastable bound states and even the states
below chemical potential. In that case the total amount of the electrons with
"wrong" orientation in the external magnetic field will be diminished by one
compare to the vortex without bound electron. The thermodynamic gain by the creation
of the vortex will be conserved. For real experimental conditions the cyclotron
energy is not very large compare to Coulomb energy and g-factor is also not
very small and therefore these statements need computer check.

\section{Hopf invariant in the action}

Besides the mapping degree $Q$ there is also topological Hopf invariant
 corresponding to the linking number of the lines with constant values of
$\vec n(\vec r,t)$ for nonstationary matrix $U$
(see e.g.[13]). This invariant must be contained in the action with the
coefficient determining the statistics of vortices  or more precise the phase
change when they are transposed [13]. This coefficient was calculated in [5]
in the model of projected wave functions. However this calculation contains
some additional assumptions and was questioned in the discussion [6].
The authors of [6] reference to their previously made semiclassical
calculation which use the frequency and the momentum as a good quantum numbers
for Green function calculation and cite the proper result. In this section
we make a calculation of this coefficient directly in the limit of large
magnetic field without additional assumptions. For that one needs to find the
action in the third order of $\Omega$ including only the terms containing
two space and one time component of $\Omega^l$ .

Third order perturbation terms in the action are
\begin{equation}
\label{h1}
S=iTr[\frac{1}{2}H_1G_0H_1G_0+(H_1G_0H_2G_0)+
\frac{1}{3}H_1G_0H_1G_0H_1G_0]
\end{equation}
The terms up to the second order were calculated in the previous section.
However time derivatives were not included in these terms what is essential
in the third order calculations. We shall calculate the
various third order terms one by one.

We begin with the second order in $H_1$ term containing $\Omega_t^l$ omited
in the previous section assuming $\Omega_t\sim\vec \Omega^2$. After an easy
calculation we get
\begin{displaymath}
\nonumber
S_2^1=\frac{i}{m}\int Sp\sigma_lg_s(\omega)\sigma_{l'}{g_s'}(\omega)e^{i\omega\delta}
\Omega_t^l(r,t)\Phi_{sp}(\vec r)\Phi^{*}_{sp}(\vec r_1)\times\\
\end{displaymath}
\begin{equation}
\nonumber
(\Omega_{+}^{l'}(\vec r_1,t)\pi^{-}+\Omega_{-}^{l'}(\vec r_1,t)\pi^{+})\Phi_{s'p'}(\vec r_1)
\Phi_{s'p'}^{*}(\vec r) d^2rd^2r'dt\frac{d\omega}{2\pi}
\end{equation}
 With the needed accuracy there are one term with $s=s'=0$ and two terms with
$s=1,s'=0;s=0,s'=1$. Introducing the new integration variables
$\vec R=(\vec r+\vec r_1)/2,\vec \rho=\vec r_1-\vec r$ and expanding
 $\Omega_t(\vec r),\vec \Omega(\vec r_1)$ in
$\vec\rho$ up to the first order we obtain for the first term
\begin{displaymath}
S^1_{200}=\frac{i\sqrt2}{2m}\int Sp \sigma_lg_0(\omega)\sigma_{l'}g_0(\omega)
e^{i\omega\delta}(\Omega_t^l(\vec R)\frac{\partial\Omega_{-}^{l'}}{\partial R_k}-
\Omega_{-}^{l'}(\vec R)\frac{\partial\Omega_t^l}{\partial R_k})R_{00}(-\vec \rho)
R_{10}(\vec \rho)\rho_k d^2\rho d^2Rdt\frac{d\omega}{2\pi}
\end{displaymath}
Using expressions (\ref{g2}), (\ref{g3}) and performing the proper integrals in
$\omega$ and $\vec \rho$ we have
\begin{equation}
\label{h2}
S_{200}^1=\frac{1}{4m\gamma}\int \left[i\Omega_t^l div \vec \Omega^l-
i(\vec \Omega_l\nabla)\Omega_t^l-\Omega^l_t\nabla\vec\times \Omega^l-
\vec \Omega^l\times\nabla\Omega^l_t\right]\frac{d^2rdt}{2\pi}
\end{equation}
The terms with $s=0,s'=1;s=1,s'=0$ are calculated in the same way
\begin{displaymath}
S_{201}^1=\frac{1}{2}\int \left[\Omega_t^l\nabla\times \Omega^l+\vec \Omega^l\times\nabla\Omega^l_t\right]\frac{d^2rdt}{2\pi} \\
\end{displaymath}
\begin{equation}
\label{h3}
+\frac{1}{4}Sp \sigma_l\sigma_{l'}\sigma_z\int \left[-i\Omega_t^ldiv\vec \Omega^{l'}
+i\vec \Omega^{l'}\nabla\Omega^l_t\right]\frac{d^2rdt}{2\pi}
\end{equation}

There is also the term with two $\vec \Omega^l$ considered in the previous
section not taking into account time derivatives. We can proceed quite similar
introducing the new integration variables  $T=(t+t')/2,\tau=t'-t$ and expanding in
$\tau$. With the required accuracy it is sufficient to consider omly  the terms
with $s=1,s'=0$ and $s=0,s'=1$ others contain extra derivatives.
Omiting the second order term without time derivatives we get
\begin{displaymath}
S^2_{2}=\frac{i}{m^2}\int Sp \sigma_lg_0(\omega)\sigma_{l'}g_1(\omega')e^{i\omega\delta}
\tau\exp{i(\omega-\omega')\tau}[\Omega^l_{-}\frac{\partial\Omega_{+}^{l'}}{\partial T}-
\Omega_{+}^{l'}\frac{\partial\Omega_{-}^l}{\partial T}]\frac{d\omega d\omega'}
{(2\pi)^2}d\tau \frac{dT d^2r}{2\pi}
\end{displaymath}
Representig $\tau$ as the derivative of the proper exponent in the integrand
and integrating by parts we have
\begin{displaymath}
S^2_{2}=-\frac{1}{2}\int \vec \Omega^l\times\frac{\partial\vec \Omega^l}{\partial t}
\frac{d^2r dt}{2\pi}-\frac{i}{4}Sp \sigma_{l'}\sigma_l\sigma_z\int
(\vec \Omega^l\frac{\partial\vec \Omega^{l'}}{\partial t}-\vec \Omega^{l'}
\frac{\partial\vec \Omega^l}{\partial t})\frac{d^2r dt}{2\pi}
\end{displaymath}

In order to compare various terms in action they must be expressed in
some standard way through $\Omega^l$ and it's derivatives. Using the identity
 $\partial_t\Omega_k^l-\partial_k\Omega_t^l=2e^{ljm}\Omega_t^j\Omega_k^m$
where $e^{ljm}$ is antisymmetrical tensor it is easy to transform the previous
expression to the following form
\begin{equation}
\label{h4}
S^2_{2}=-\frac{1}{2}\int \left(\vec \Omega^l\times\nabla\Omega_t^l+
e^{ljm}\vec \Omega^l\times\vec \Omega^m\Omega_t^j\right)\frac{d^2r dt}{2\pi}-
\end{equation}
\begin{displaymath}
\frac{i}{2}Sp \sigma_{l'}\sigma_l\sigma_z\int \left[\vec \Omega_l\nabla\Omega_t^{l'}
+2e^{l'jm}\vec \Omega^l\vec \Omega^m\Omega^j_t\right]\frac{d^2r dt}{2\pi}
\end{displaymath}

Among the third order terms the simplest is containing $H_2$ and $H_1$
\begin{equation}
\nonumber
S_{21}=\frac{i}{2m}Sp \int [(\vec \Omega^l)^2-idiv \vec \Omega^l\sigma_l]\Omega_t^{l'}
g_0(\omega)\sigma_{l'}g_0(\omega)e^{i\omega\delta}\frac{d\omega}{2\pi}\frac{d^2r dt}{2\pi}
\end{equation}
The other terms contain extra derivative. Only the term with the divergence
gives nonvanishing contribution
\begin{equation}
\label{h5}
S_{21}=-\frac{i}{2m\gamma}\sum_{l\ne z}\int \Omega_t^ldiv \vec \Omega^l\frac{d^2r dt}{2\pi}
\end{equation}

The last remaining third order terms contain only $H_1$ in the combination with
two space and one time $\Omega^l$
\begin{displaymath}
\nonumber
S_3=\frac{i}{m^2}\int \Omega_t^l\Phi_{sp}(\vec r)\Phi_{sp}^{*}(\vec r_1)
(\Omega_{+}^{l_1}\pi^{-}+\Omega_{-}^{l_1}\pi^{+})\Phi_{s_1p_1}(\vec r_1)\Phi_{s_1p_1}^{*}(\vec r_2)\\
\end{displaymath}
\begin{displaymath}
\times (\Omega_{+}^{l_2}\pi^{-}+\Omega_{-}^{l_2}\pi^{+})\Phi_{s_2p_2}(\vec r_2)\Phi_{s_2p_2}^{*}(\vec r)
d^2rd^2r_1d^2r_2dt\int Sp \sigma_lg_s\sigma_{l_1}g_{s_1}\sigma_{l_2}g_{s2}e^{i\omega\delta}\frac{d\omega}{2\pi}
\end{displaymath}
Only the terms with $s=s_2=0;s_1=1$ and $s=s_2=1;s_1=0$ gives the needed third
order and no derivatives of $\Omega$ must be taken into account.

The calculations in the first case are very similar to the calculation of
(\ref{h2}),(\ref{h5}) for double pole in the production of
$g_0$. Using the standard form in terms of $\Omega_t^l$ and $\vec \Omega_l$
we get
\begin{displaymath}
\nonumber
S_3^0=-\frac{i}{m\gamma}\sum_{l\ne z} Sp\sigma_l\sigma_{l_1}\sigma_{l_2}
\int \Omega_t^l\vec\Omega^{l_1}\times\vec \Omega^{l_2}\frac{d^2rdt}{2\pi}+\\
\end{displaymath}
\begin{equation}
\label{h6}
\frac{i}{4}Sp \sigma_z\sigma_{l_1}\sigma_{l_2}\int \Omega_t^z\vec \Omega^{l_1}\vec \Omega^{l_2}\frac{d^2rdt}{2\pi}+
\frac{1}{2}\sum_{l\ne z}\int \Omega_t^z(\vec \Omega^l)^2\frac{d^2rdt}{2\pi}
\end{equation}

The calculation of the case $s=s_2=1;s_1=0$ is standard and gives
\begin{equation}
\label{h7}
S_3^1=-\frac{1}{2}\int \Omega_t^z(\vec \Omega^z)^2\frac{d^2rdt}{2\pi}+
\frac{1}{2}\sum_{l\ne z}\int [\Omega_t^z(\vec \Omega^l)^2-\Omega_t^l\vec \Omega^z\vec\Omega^l]\frac{d^2rdt}{2\pi}
+\frac{i}{4}Sp \sigma_l\sigma_{l_1}\sigma_{l_2}\int \Omega_t^l\vec\Omega^{l_1}\times\vec\Omega^{l_2}
\frac{d^2rdt}{2\pi}
\end{equation}

The final expression for the part of the action containing the Hopf invariant is
obtained by the summation of all found terms (\ref{h7}), (\ref{h6}), (\ref{h5}),
(\ref{h4}), (\ref{h3}), (\ref{h2}). Some points in this summation must be
explained. The terms with the factor $1/\gamma$ in $S_3^0, S_2^0, S_{21}$
combine to the total derivative
\begin{displaymath}
\nonumber
\frac{1}{4m\gamma}\sum_{l\ne z}\int [-idiv (\Omega_t^l\vec \Omega^l)-
\nabla\times(\Omega_t^l\vec \Omega^l)]\frac{d^2rdt}{2\pi}=0
\end{displaymath}
because $\Omega^l$ for $l\ne z$ tends exponentially to zero at large distances.
In the same way the term containing the other total derivative
$$
\frac{-i}{4}Sp \sigma_l\sigma_m\sigma_z\int [\Omega_t^ldiv \vec \Omega^m+
(\vec \Omega^m\nabla)\Omega_t^l]\frac{d^2rdt}{2\pi}=0
$$
vanish.

The terms with $\Omega_t^z(\vec \Omega^l)^2$ and $\Omega_t^l(\vec \Omega^z\vec \Omega^l)$
combine also to zero after some algebraic manipulations with $S^2_2$. Finally
only antisymmetric terms in $S_3$ are retained
\begin{displaymath}
\nonumber
S_3=\frac{i}{4}Sp \sigma_l\sigma_{l_1}\sigma_{l_2}\int \Omega_t^l\vec \Omega^{l_1}
\times\vec \Omega^{l_2}\frac{d^2rdt}{2\pi}+\frac{1}{2}\int \Omega_t^l \nabla\times\vec \Omega^l\frac{d^2rdt}{2\pi}-
\int e^{ljm}\Omega_t^j\vec \Omega^l\times\vec \Omega^m\frac{d^2rdt}{2\pi}
\end{displaymath}
Calculating the trace and using the identity for $\nabla\times \vec \Omega^l$ we get
\begin{equation}
\label{h8}
S_3=e^{ljm}\int \Omega_t^l\vec \Omega^j\times\Omega^m\frac{d^2rdt}{2\pi}
\end{equation}
According to [13],[12] the Hopf invariant can be expressed in terms of $\Omega$
\begin{displaymath}
\nonumber
H=\frac{1}{2\pi^2}\int e^{ljm}\Omega_t^l\vec \Omega^j\times\vec\Omega^m d^2rdt
\end{displaymath}
Therefore the Hopfian term in vortex action is
\begin{equation}
\label{h9}
S^H=\pi H
\end{equation}.
This result coincide with those in [6] . According to the accepted point of
view that means [13] the fermionic nature of vortices-skyrmions.

We must remain that the Hopf invariant $H$ is integer for the mapping of the sphere
$S_3$ on the sphere $S_2$ that requires in our case vanishing of $\Omega$ at
large $\vec r,t$ i.e. $Q=0$. If the vortices are present at any time and
$Q\ne 0$ then the topological invariant corresponding to the linking number
changes a bit and is contained in some part of $H$ [15],[16].However the found
relation between $S$ and $H$ holds on as well as the statement about fermionic
nature of vortices.

\section{Conclusions}

In this work the theory of nonsingular vortices-skyrmions was considered
for 2d electron systems in a strong magnetic field. It was assumed that
g-factor is sufficiently small to justify the perturbation theory in the
derivatives of the rotation matrix. No other assumptions beside the large
 value of cyclotron energy $\hbar\omega_c$ compared to Coulomb energy $e^2/l_H$
was used. It was shown that currently used approximation of the single Landau level
projected wave functions is not sufficient for the adequate description.
Taking into account the nearest Landau levels is essential and changes the
thermodynamic energy for vortex-skyrmion formation diminishing it's value by
$-\hbar\omega_c/2$ large by assumption. This fact must lead to the spontaneous
creation of vortices with positive degree of mapping near odd Landau level
fillings. Simple physical picture emerges for small enough g-factor providing
validity of the perturbation theory : inside the vortex core exists an additional
effective magnetic field with the total flux containing the number of the flux
quanta equal to the degree of the mapping. The lowest Landau spin sublevel in the total
(external plus effective)local field is fully filled and is separated by the
exchange gap from the higher energy states. The total number of electrons
changes because of the change of their density coinciding with the local
density of states. As aconsquence the electrical charge of the core emerges
which equals t<< that of electrons expelled from the core in the number of flux
quanta for a positive degree of mapping.

There is a possibility to have bound electron states in the positive charge of
the core.The position of the proper energy is defined by negative Coulomb
energy and positive exchange energy not depending on the core.
Under made assumptions it is above chemical potential. In real experimental
situation this statement must be checked numerically.

The topological Hopf term in the vortex-skyrmion action is found. The calculated
coefficient before the Hopf invariant coincides with semiclassical result [6].
Therefore the vortices are fermionic-like. The properties of
vortices-skyrmions are close to "composite" fermions introduced phenomenologically.
in [14].

Authors express their gratitude to G.E.Volovik for valuable discussions.

The research described in this publication was made in part due to award
 RP1-273 US Civilian Research and Development Foundation
for Independent States of Former Sovjet Union. This work supported partially
also by RFFI 95-02-05883 and the program "Statistical Physics" of the Russian
Science Ministry.


\end{document}